\documentclass[showpacs,preprintnumbers,amsmath,amssymb]{revtex4}
\usepackage{dcolumn}

\newcommand{\beq}{\begin{equation}}
\newcommand{\eeq}{\end{equation}}

\newcommand{\eq}[1]{eq.(\ref{#1})}

\begin{document}
\title{Is the Luttinger liquid a new state of matter? }
\author{V. V. Afonin}
\affiliation{Solid State Physics Division, A.F.Ioffe  Institute,
194021 St.Petersburg, Russia}
\author{V. Yu. Petrov}
\affiliation{Theory Division, St.Petersburg Nuclear Physics
Institute, 188300, St.Petersburg, Russia}
\begin{abstract}

We are demonstrating that the Luttinger model with short range
interaction can be treated as a type of Fermi liquid. In line with the
main dogma of Landau's theory one can define a fermion excitation
renormalized by interaction  and show that in  terms of these
fermions any excited state of the system is described by free
particles. The fermions are a mixture of renormalized right and
left electrons. The electric charge and chirality of the Landau
quasi-particle is discussed.
\end{abstract}
\pacs{71.10.Hf, 73.63.Fg}
\maketitle

\section{Introduction}
\label{I}

The theory of one-dimensional interacting fermions (1D system) has
been developed for more than five decades. First time one
considered the problem as a simplification of a real 3D task
\cite{Sh} but later it was realized that 1D problems are
interesting by themselves (especially for the solid state physics
where 1D systems are accessible to a direct measurements). One of
the most essential achievements  in 1D theory without back
scattering was a demonstration of the fact that long wave
excitations can be expressed in terms of non-interacting bosons
(Tomonaga \cite{T} and Luttinger \cite{L}). In explicit form these
expressions were presented by Mattis and Lieb \cite{ML}. In the
subsequent papers this approach was called "bosonization".
Bosonization procedure allows one to proceed from strong
interacting 1D electrons to the non-interacting Bose-particles. In
other words, the Fermi-type excitations  disappear from the
theory. It was one of the immediate reasons which led to the
conclusion that Fermi liquid theory breaks down in 1D systems. The
other one is the form of a Green functions of right (or left)
electrons (see e.g. the review paper \cite{Schu}). Instead of
poles they have only a branch cut, i.e.  do not correspond to a
free fermion excitation. As a result it is commonly believed that
the system of 1D interacting electrons is a new state of matter
which is not described by the Fermi liquid theory
\cite{Voit}("non-Fermi liquid state"). "Luttinger liquids (LL) are
non-Fermi liquids: Landau quasi-particles are not elementary
excitations of the LL and as a consequence the electron Green's
function shows no quasiparticle pole..." (citation is taken from
the paper \cite{Ph}.) We cannot agree with such argumentation.
From our point of view, the form of renormalized electron Green
function demonstrates only that the electrons themselves should
not be considered as candidates for Landau quasi-particles.
Indeed, the renormalized electrons are weak interacting Landau
quasi-particles in a 3D normal metal where the Coulomb interaction
is sufficiently weak \cite{AAA}. However, the Landau theory does
not require that quasi-particle should coincide with renormalized
electron in all cases. According to Landau \cite{Lan}, weak
interacting quasi-particles should exist but their origin is not
specified. Notice, in a wide sense the term "Fermi-type liquid"
was already used in \cite{AAA} for the normal excitations in 3D
superconductor. Here the weakly interacting excitation is a
superposition of the electron and hole with opposite momenta and
spins \cite{V}; it does not coincide with a renormalized electron.
Late on, as a consequence of the Fermi liquid behavior of normal
excitations, one has understood that a long range excitation in 3D
superconductors can be described by the classical kinetic equation
(similar to the Boltzmann equation) \cite{Koz}. This approach
appears  extremely successful and enables one to solve numerous
problems in superconductivity.

The strongest interaction in the Luttinger model is the
interaction between right and left electrons. Hence one can expect
that well-defined quasi-particle is a mixture of right and left
electrons (c.f. with 3D superconductivity) rather than simple
renormalized electron. Reason of the assumption is in absolute
necessity to include a strong interaction (changing the ground
state of a system) in the initial approximation in order to define
a new exact ground state. One cannot consider such interaction
perturbatively. Excited states have to be built over the new {\em
stable } ground state. Only such excitations  have a chance to be
well-defined Landau quasiparticles.  Such approach was used by
Carmelo and Ovchinnicov for the Hubbard model \cite{J1} and
Carmelo and Horsch \cite{J3a} for the Hubbard chain in a magnetic
field. Later \cite{J3}, it was shown that low-temperature
thermodynamics of the Hubbard model  can be explained in terms of
weak interacting quasiparticles that are defined over interacting
ground state.  This method  seems to be more acceptable than
examination of a form of renormalized electron Green function
(as used here the last is an attempt to build a well-defined
quasiparticle over filled 1D Fermi sphere unstable relative to
infinitesimal interaction).

We will see below that the analogy with 3D superconductivity is
justified and one can introduce noninteracting quasi-particles of
such type for Luttinger model with short range interaction. This
does not mean that that bosonization procedure is invalid. The
system can be described equally well in both representations,
either of free bosons or free fermions.  In fact, this should be
expected from the very beginning. Indeed, the equivalence of boson
and fermion representation for the non-interacting system is
well-known. Let us recall the physical reasons of this phenomenon.
The fermion-boson equivalence is due to the linearity of the
Hamiltonian and one-dimensionality of the system. In such a system
it is not possible to distinguish the wave packet composed of a
number of the particles from an elementary particle. In other
words, there is no way to judge, whether given excitation is
elementary or consists of many elementary excitations. Usually one
can distinguish one particle state if the total energy of the
state is a function only of a total momentum of this state (no
internal momenta). However, in the case under consideration this
is always true for any number of particles. Mathematically this
means that in free one-dimensional system with linear spectrum one
can construct a stable packet with virtually any transformational
properties and call it elementary excitation. What really matters
is the influence of the interaction on the stability of a packet.
However it is well-known the effect of short-range interaction in
the Luttinger model reduces only to the renormalization of the
excitation velocity while the spectrum remains linear. Therefore a
wave packet and a single particle do not differ even in the model
with interaction. Thus, free fermion and boson descriptions should
be possible at the same time. Moreover, even in the model with
non-local (but decreasing with distance) e-e interaction, the long
wave excitations can be described in fermion language in the
region where wavelength is larger than the characteristic size of
the interaction. On the other hand, if e-e interaction does not
decrease with distance, it is evident that fermion states cannot
exist. One calls  this as a confinement of electrons. Well-known
example is  the Schwinger \cite{Sh} model where the e-e potential
is a linear ({\em increasing!}) function of the distance. In this
model  the energy spectrum has a gap and our reasoning is not
valid. Only boson description is possible in this model.

Can one call the 1D system with short range interaction that is
described in terms of  noninteracting fermions as a Landau-type
liquid? We believe the answer is positive. According to the main
principles of the Landau theory \cite{AAA} one should be able to
define a quasi-particles and show that in terms of new fermions
the excitations of the system are described by free
Hamiltonian. This is exactly the case. As regards the observed values,
one can rewrite it in the terms of new quasi-particles.

As appears from the above, it is possible to construct packets
with virtually any transformational properties and quantum numbers
\cite{Ph}(electric charge, chirality, fermion number, etc) in the
interacting theory. Most of them will remain strong interacting.
Nevertheless, they can be stable too. However, only weak
interacting packets can be considered as final excitations of the
interacting theory. In the Luttinger liquid with a short range
interaction these are either bosons or fermions with definite
quantum numbers. What representation to use is, in a sense, a
matter of convenience so far as some additional interaction
removes degeneracy of one particle states and the states
consisting  of many elementary excitations. We know that in the
Schwinger model the degeneracy is removed in favor of bosons. This
means that fermions are confined and it is due to linear growth of
the inter-electron potential. We see no reason for existence of
such phenomenon in the ordinary non-relativistic 1D models with
decreasing interaction and believe that long range excitations of
the system can be described as a Fermi liquid of Landau type.

Finally, let us compare the conception asserting that the
Luttinger liquid is a new non-Fermi liquid state  and the
viewpoint offered in the present paper. The first was formulated
e.g. in \cite{Ar} as a motivation of the work. "The concept of
Luttinger liquid is an alternative to the Fermi liquid elaborated
for one-dimensional electronic systems. It was found that in 1D
electronic systems the Fermi-liquid picture breaks down even in
case of arbitrary weak interaction. Single-electron quasiparticles
cannot exist in 1D metals, and electrons form the Luttinger liquid
in which the only low energy excitations turn out to be charge and
spin collective modes with the soundlike spectrum..." According to
this viewpoint, the Luttinger liquid segregates from other
physical systems and has nothing in common with other 3D electron
systems. On the contrary, we believe that the excitations in the
Luttinger liquid with a short range interaction can be described
as well as a Fermi-liquid state. In this regard the Luttinger
liquid is similar to 3D systems. (An immediate analogy is the
normal excitations in 3D systems with a long range order, e.g. in
superconductors.) At the same time, a significant difference
between Luttinger liquid and 3D system exists. First, because of
one-dimensionality, in the Luttinger liquid with a short range
interaction the free fermion and free boson representations
coexist. Secondly, owing to strong e-e interaction relation
between Landau quasi-particles and renormalized right and left
electrons (similar to Bogolubov-Valantin transformation \cite{V}
in 3D superconductivity) is non-linear, etc. However, there are
details. In general excited states of the systems qualitatively
resemble. (Existence common features of 1D Hubbard chain and 3D
Fermi liquids has been indicated in  \cite{J2}.)

At the same time resemblance does not mean identity. In the LL
model the electron fluxes differ from quasiparticle ones. It is a
usual situation for a system with a long range order because in
this case a quasiparticle consists of a mixture of different
electron states. So, if one expresses  one flux through other
extra factors appear. (In 3D superconductivity one calls them
"coherence factors.") They change the electron transport
coefficients (defined by normal excitations) in comparison with
ordinary expressions for the conventional Landau liquid. For
Luttinger liquid the difference is even more essential because
relation between creation operator of an electron and the
quasiparticle operators is nonlinear (see below). So, physical
properties of the system can differ essentially  from a
conventional Fermi system. However, the crucial point for the
Fermi liquid theory is existence  well-defined  quasiparticles and
this is the case. In this regard one may treat  excitations of the
Luttinger liquid (as well as normal excitations in other systems
with long range order) as a generalized Landau Fermi liquid. The
main message of the paper is: "For a short range e-e interaction
the conceptions of the Luttinger and Landau liquids do not
contradict one another but coexist."

The last point has to be considered in the paper is the relation
between ferminization of the Luttinger liquid and the
Dzyaloshinsii-Larkin theorem \cite{Lar}. According the theorem
Random Phase Approximation approach (RPA) is exact for the model
with short range interaction.  We discuss the problem in the
Appendix \ref{lar}.

\section{Fermion representation of the excitation.}
\label{G}
\subsection{Spinless fermion}
\label{1}

In order to introduce a notation and  discuss a point which is
crucial for further discussion  we begin with a brief review
of the well-known bosonization procedure.  As usual, we  divide
the electron wave function $\hat\Psi (x)$ into wave functions of
the left- and right- particles ( $\hat\Psi_{L,R}$
respectively):
\begin{equation}
\hat \Psi \left( x\right)=\exp\left( ip_f x\right)\hat \Psi _R\left( x \right)
+\exp\left( -ip_fx\right)\hat \Psi_L \left( x \right)
\label{psi} ,
\end{equation}
 here $p_f$ is the Fermi momentum. Throughout the paper we are
 interested in a model with short-range electron-electron interaction:
\beq
H=\int dx \left[ \hat \Psi^{\dag}_R \left( x
\right) v_f\left( -i \partial_x \right) \hat \Psi_R \left( x  \right) +
 \hat \Psi^{\dag}_L \left( x \right) v_f i\partial_x
\hat \Psi_L \left( x  \right) \right] + \int dx
\hat\varrho\left( x  \right)V_0 \hat\varrho\left( x \right),
\label{g5}
\eeq
where $\hat\varrho\left( x  \right)$ is the operator of the
total electron density:
$$
\hat\varrho\left(  x  \right)=\hat\varrho_R\left( x  \right)+
\hat\varrho_L\left( x  \right).
$$

Note that the densities of the left- and right-particles,
$\hat\varrho_{R,L}$, obey the well-known Schwinger anomaly \cite{Sh}:
\beq
 \left[ \hat\varrho_{R,L}\left( x
\right),\hat\varrho_{R,L}\left( y \right)\right]= \pm
\frac{i}{2\pi}\frac{\partial}{\partial x} \delta\left( x-y \right).
\label{g8}
\eeq
The last equation is a starting point for bosonization procedure \cite{ML}.
One introduce two  Bose fields corresponding to the left- and right-electron densities:
\begin{equation}
\hat C_{R,L}\left( p\right) = \sqrt {\frac{2\pi}{ p }}\int dx \exp
(\mp ipx )\hat\varrho_{R,L}\left( x  \right) .
\label{bos}\end{equation}
So, fields $\hat C_{R,L}, \hat C_{R,L}^{\dag}$ commute
in the way usual for Bose-particles
(in Eqs. (\ref{bos}),(\ref{R}) all $p>0$). The following
transformation
\begin{equation}
\hat C\left( p\right)= \cosh\theta \hat C_{R}\left( p\right)+
\sinh\theta\hat C^{\dag}_{L}
\left( p\right)\\
\label{R}
\end{equation}
$$
\hat C^{\dag}\left( -p\right)= \cosh\theta \hat C_{L}^{\dag}\left(
p\right)+ \sinh\theta\hat C_{R} \left( p\right)
$$
allows to diagonalize the Hamiltonian  if $\theta$ is chosen in such a way that
$ \sinh2\theta = V_0/2\pi v_f^c $
(which retains the commutator of the operators):
\beq
H=v_f^c \int^{\infty}_0 \frac{d p}{2\pi}p\left( \hat
C^{\dag}\left( p\right) \hat C \left( p\right)+ \hat
C^{\dag}\left( -p\right)\hat C\left( -p \right) \right).
\label{diag}
\eeq
 Here
$$ v_f^c=v_f \sqrt{ 1 +\frac{V_0 }{\pi v_f}}$$
is the renormalized  Fermi velocity (see, for example \cite{M}).
It is convenient for as to use the bosonization scheme operating
with particles $\hat C_{L,R}$  (see \cite{P2}) rather than a
scheme with a total electron density and momentum canonically
conjugate to it. It is possible to present $\Psi$ in terms of
boson operators as follows:
\begin{equation}
\Psi^{\dag}_{R,L}\left( x \right) = \exp\left(
{\cal{A}}^{\dag}_{R,L}\left( x  \right)
\right)\frac{\hat\sigma^{\dag}_{R,L}} {\sqrt{L}} \exp\left( -
{\cal{A}}_{R,L}\left( x  \right) \right) . \label{bos1}
\end{equation}
Here
$$
{\cal{A}}^{\dag}_{R,L}\left( x  \right)=
\int_{2\pi/L}^{\infty}\frac{dp}{2\pi} \exp{(\mp
ipx)}\sqrt{\frac{2\pi}{ p}}\hat C^{\dag }_{R,L}(p) ,
$$
while $\hat\sigma$ is the operator similar to the ladder operator
introduced by Haldane \cite{Hold}. In order to have correct
commutation rules for the $\Psi_{R,L}$ one should require:
$\hat\sigma^{\dag}_{R,L}\hat\sigma_{R,L}=1 $, $\{
\hat\sigma_{R,L},\hat\sigma_{L,R}  \}=0 $. Besides, one can see
that $\hat\sigma$ and $\hat\sigma^{\dag}$ commute with  $ \hat
C_{L,R}$.

However, mathematically the exact Hamiltonian (Eq.\ref{diag})
differs from the boson Hamiltonian originated of free electrons
only by the replacements $v_f^c \to v_f$ and $\hat C \left(
p\right)\to \hat C_{R,L} \left( p\right)$. So, if one defines {\em
fermion} operators ($\hat\chi$)  coinciding with original $R$ and
$L$-electrons in the limit $V_0\to 0$:
\begin{equation}
\hat\chi^{\dag}_{\pm}\left( x \right) = \exp\left(
A^{\dag}_{\pm}\left( x  \right) \right)\frac{\hat\sigma^{\dag}_{\pm}}{
\sqrt{L}} \exp\left( - A_{\pm}\left( x  \right) \right)
\label{ferm1}
\end{equation}
with
$$
A^{\dag}_{\pm}\left( x  \right)= \int_{2\pi/L}^{\infty}\frac{dp}{2\pi}
\exp{(\mp ipx)}\sqrt{\frac{2\pi}{ p}}\hat C^{\dag }(\pm p) ,
$$
then one can see directly
\begin{equation}
\{ \hat \chi^{\dag}_{\pm}\left( x  \right) ,\hat \chi_{\pm}\left(
x_1  \right)\} =  \delta (x-x_1) \label{ferm2}
\end{equation}
(for $|x-x_1|\ll L$) and show that in terms of the
$\chi$-particles our system is described by the free Hamiltonian
without interaction. (In fact, this is the transformation an
electron kinetic energy from fermion to boson representation with
replacements noted above. See, for example, \cite{ML}.)  Hence,
the fermions are nothing more than original electrons excitation
moving together with ground state polarization: \beq H =v_f^c\int
dx  \left( \hat\chi^{\dag}_+\left( x  \right)\left( -i\partial
_x\right) \hat\chi_+\left( x  \right) + \hat\chi^{\dag}_-\left( x
\right)\left( i\partial _x\right) \hat\chi_-\left( x
\right)\right). \label{diagf} \eeq Formally it is shown in
Appendix \ref{f}, there one can see that it is right only for the
short-range interaction ($v_f^c$ does not depend on $p$).
Otherwise, four - fermions terms should be added to
Eq.(\ref{diagf})  \cite{Gl}.

Both Hamiltonians,  Eq. (\ref{diag}) and Eq.(\ref{diagf}),
describe all excited states of the system but in different
representations. It is important that the ground state wave
function which is well-known in terms of bosons is, in fact, the
vacuum state for excitations in both representations. In order to
prove that the state with lowest energy (see, for
example,\cite{Hold})
\begin{equation}
| GS_0 > = N \exp \left( -\int_{0}^{\infty}\frac{dp}{2\pi} \tanh
\theta \hat C^{\dag }_L \left( p  \right) \hat C^{\dag }_R \left(
p \right)\right)| F > \label{bos6} \eeq is the vacuum state for
renormalized field $\hat\chi$ one should proceed from whole
fermion fields $\hat\chi$ to the particle - hole representation ($
{\frak a}_{\pm}\left( x  \right), {\frak b}_{\pm}\left( x
\right)$) :
\begin{equation}
\hat \chi_{\pm} \left( x  \right)= \int \limits_0^{\infty}
\frac{dp}{2\pi} \left( \exp\left( \pm ipx \right)\hat {\frak
a}_{\pm}\left( p  \right) +
\exp\left( \mp ipx \right)\hat {\frak b}_{\pm}^{\dag}\left( p  \right)\right)=\\
\hat {\frak a}_{\pm}\left( x  \right) + \hat {\frak
b}_{\pm}^{\dag}\left(x \right). \label{psi2}
\end{equation}
($N$ in Eq.(\ref{bos6}) is the normalization coefficient.)
So, the electron part can be extracted from  $\hat\chi $ in the
following way:
\begin{equation}
{\frak a}_{+}\left( y\right) = \frac{1}{2\pi i}\int
dx\frac{\hat\chi_+\left( x  \right)}{x-y -i\delta} .
\label{o1}
\end{equation}
Hence
\begin{equation}
{\frak a}_{+}\left( y\right)| GS_0 >=\hat\sigma_+\frac{1}{2\pi i\sqrt
{ L}}\int\frac{dx} {x-y -i\delta}\exp{\left(-\int_{2\pi/L}^{\infty}
\frac{dp}{2\pi} \exp{(- ipx)}\sqrt{\frac{2\pi}{ p}}\hat C^{\dag }(
p)\right)}| GS_0 > = 0 , \label{o2}
\end{equation}
because $ \hat C(p) |GS_0 >=0$ and the integrand of Eq.(\ref{o2})
has no singularities in lower semiplane. One can prove the same
property for the hole part as well. So, in spite of the fact that
the ground state Eq.(\ref{bos6}) is not empty, the $\hat {\frak
a}-$particles are not present there. That is, one can consider the
particles $\hat {\frak a}, \hat {\frak b}$  as excitations over
the ground state Eq.(\ref{bos6}). (This requirement is necessary
because a ground state is the state with smallest energy, i.e. it
is the state without excitations.) Notice, this claim does not
hold for an $R(L)-$electron because $\hat C_{R,L}(p) |GS_0
>\ne 0$. It  means that either the R(L)- electron cannot be
considered as an excitation of LL or the ground state
Eq.(\ref{bos6}) is defined incorrectly. We believe this is the
first case.

For larger temperatures $ 2\pi v_f^c/L\gg T\gg 2\pi v_f/L =
T_{chiral}$ the states with different chiralities are degenerate
(we have in mind a strong repulsive e-e interaction here),
because the $ 2\pi v_f / L $ is the characteristic energy
difference between the states. (Note, the temperature region $2\pi
v_f^c/L\gg T$ is the region where power - low correlators  exist.
At the same time, the state Eq.(\ref{bos6}) is the state with zero
chirality.) So, the real ground state for the considered
temperature region is a mixture of the states with all chiralities
and, as a result, it is the state with broken chiral symmetry:
\begin{equation}
|\theta>= \sum_{-\infty}^{\infty}\exp{\left(in\theta\right)}|GS_n>
, \label{bos7}
\end{equation}
here
$$|GS_n>=\left(  \hat\sigma_L\hat\sigma^{\dag}_R \right)^n|GS_0> \phantom {.a} for \phantom{..}
n > 0; \phantom {..} |GS_n>=\left(  \hat\sigma_R
\hat\sigma^{\dag}_L\right)^n|GS_0> \phantom{. .} for {\phantom .}n
< 0 .  $$
(See \cite{A} for detailed discussion.) It is obvious
that  this state is the vacuum state for both representations too.
So, as it should be, well-defined quasi-particle excitations can
be obtained  only over the stable exact ground state, and not over
a filled 1D Fermi sphere. The last state is unstable relative to
infinitesimal e-e interaction.

Let us discuss such characteristics of the renormalized fermion
$\hat\chi $ as electric charge and chirality.  We assign chirality
$ +1 $ to a right electron, and a left hole
and $ -1 $ to their counterparts. The corresponding "density" is
$\hat\Sigma \left( x \right) = \hat\varrho_R\left( x \right) -
\hat\varrho_L\left( x \right)$. So, the last charge is the charge
determining contribution to the electric current from a
renormalized fermion.

For a system of weak interacting electrons the above question is
trivial. In a conventional Fermi liquid the electric charge of a
quasi-particle equals to the bare electron one. Here we imply, as
usual, that the screening length is a  macroscopic one. Then one
can measure a charge at the distances  smaller  than the screening
length. In a system with a strong coupling, screening, total or
partial, can be realized at the distances smaller than a minimal
scale of the theory. A well-known example is the screening of a
bare electron charge in quantum electrodynamics. It is also the
case for our problem because in the main order in $V^{-1}_0$ the
electron system is polarized so strongly that the ground state
eq.(\ref{bos6},\ref{bos7}), in fact, consists of  exciton-like
neutral pairs   \cite{A} (we have in mind a repulsive interaction here).
In other words, the "screening scale" is about the transverse size of the channel.
This scale is considered to be zero by an effective 1D theory.
In this paper we will see that a similar screening takes place also for the fermion excitations.

In order to define the electric charge of a new fermion we will
use the following relations (see, for example, \cite{ch})
\begin{equation}
e_0 \left[ \hat\varrho\left( x \right) ,\hat\chi_+\left(y\right) \right]_-
=-e^*\delta\left( x-y \right) \hat\chi_+\left(y\right) .
\label{bos8}
\end{equation}
Here $e_0$ and $e^*$ are the charges of the bare electron and
exact fermion field, respectively. Similar equation is valid for
the chirality if one substitutes $\hat\Sigma$ in place of
$\hat\varrho$. During the calculation it is convenient to express
the density and chiral operators from LHS of Eq.(\ref{bos8}) in
the form:
\begin{equation}
\hat\varrho\left( x \right) =\int_{-\infty}^{\infty}\frac{dp}{2\pi}
\sqrt{\frac{|p|\gamma}{2\pi}} \left(\exp{(ipx)}C\left(p\right)
+ \exp{(-ipx)}C^{\dag}\left(p\right)\right) \label{bos9}
\end{equation}
$$
\hat\Sigma\left( x \right) =\int_{-\infty}^{\infty}\frac{dp}{2\pi}
\sqrt{\frac{|p|}{2\pi\gamma} }
sign\left(p\right) \left(\exp{(ipx)}C\left(p\right) +
\exp{(-ipx)}C^{\dag}\left( p\right)\right),
$$
$ (\gamma = v_f/v^c_f = \exp{(-2\theta)}.) $ Direct calculation gives
$$
e^*=\sqrt\gamma e_0
$$ and
\begin{equation}
\left[\hat \Sigma\left( x \right) ,\hat\chi_+\left(y\right) \right]_-=
-\frac{1}{\sqrt\gamma}\delta\left( x-y \right) \hat\chi_+\left(y\right) ,
\label{bos10}
\end{equation}
i.e. the chiral charge, $\sigma$ (corresponding to the operator
$\hat\Sigma$) equal to ${1/\sqrt \gamma}$. Therefore, exact field
$\hat\chi_+$ is not simply screening R - electron but rather a
combination of  left- and -right electrons (see relations just
below and the next section). Such a relation between the electric
charge and chirality allows one to have a finite value of the
electric current flowing through the channel (due to an
excitation), even for the strong interaction case ($\gamma \to 0
$).

In order to calculate an observable data one should establish
connections between operators $\chi$ and $\Psi$. The direct
relation $$\hat \chi_+\left( x \right) \propto
\hat\sigma_+[\hat\sigma^{\dag}_R\hat\Psi_R\left( x
\right)]^{\cosh\theta}[\hat\sigma_L\hat\Psi_L^{\dag}\left( x
\right)]^{\sinh\theta}
 $$
(and a similar one for $\hat \chi_-$) in principle gives an
opportunity to express one matrix element through another. (We do
not write out in full c-number factor here.) However, the
self-consistent equation (\ref{ferm4}) together with
Eqs.(\ref{bos}, \ref{R}) give a more convenient way for
calculation some of observed quantities:
 $$\hat\varrho_+\left( x \right)=\cosh\theta\hat\varrho_R\left( x \right)+
 \sinh\theta\hat\varrho_L\left( x \right).$$
Similar relations can be written for other bilinear
in  $\Psi_{R,L}$ operators.

\subsection{Two-component fermion.}

For a simplicity in this section we will be interested in most symmetrical
version of e-e interaction \cite{M}:
\beq H_{int}=1/2\sum_{\alpha, \beta =
1,2}\int dx \varrho_{\alpha}\left( x  \right)V_0
\varrho_{\beta}\left( x \right) .
\label{gs}
\eeq
As usual, one defines  Bose - fields $\hat C_{R,L,i}\left( p\right)$
similarly to Eq.(\ref{bos}) (with an additional spin index $i=1,2$) and
separates spin and  space variables: \beq \hat C_{R,L}\left(
p\right)=1/\sqrt{2}\left( \hat C_{R,L,1}\left( p\right) + \hat
C_{R,L,2}\left( p\right)\right); \hat S_{R,L}\left(
p\right)=1/\sqrt{2}\left( \hat C_{R,L,1}\left( p\right) - \hat
C_{R,L,2}\left( p\right)\right). \label{gs1} \eeq In the case
the spin variable separates and can be described by free
Hamiltonian with non-perturbed Fermi velocity \cite{M}. Therefore
ferminization of this part of the Hamiltonian is trivial (see
above). The spinless part of the Hamiltonian can be diagonalized
by the same transformation as previously, Eq.(\ref{R}), but with
new fields $\hat C_{R,L}\left( p\right)$ (Eq.(\ref{gs1})) and
the same rotation angle $ \exp{(-2\theta )} = v_f/V_f^c$  where
the expression for renormalized velocity should be slightly
modified:
$$ V_f^c=v_f \sqrt{ 1 +\frac{2 V_0 }{\pi v_f}}.$$
As a result, the free-fermion Hamiltonian (similar to
Eq.(\ref{diagf})) will depend on $ V_f^c $ if we define the new
Fermi-fields $\hat\Xi_{\pm}\left( x \right)$ as above:
\begin{equation}
\hat\Xi^{\dag}_{\pm}\left( x \right) = \exp\left(
A^{\dag}_{\pm}\left( x  \right) \right)\frac{\hat\sigma^{\dag}_{\pm}}{
\sqrt{L}} \exp\left( - A_{\pm}\left( x  \right) \right) .
\label{ferm1a}
\end{equation}
Let us note that the field $\hat\Xi^{\dag}_{+}$ does not turn
into the right-electron if the interaction is switched off:
$V_0\to 0$.

Had one considered the other fields $\tilde\Xi^{\dag}_{+,i}$
transferring to $\Psi_{R,i}^{\dag}$ in the limit $V_0\to 0$:
\begin{equation}
\tilde\Xi^{\dag}_{\pm ,i}\left( x \right) = \exp\left( \tilde
A^{\dag}_{\pm ,i}\left( x  \right) \right)\frac{\hat
\sigma^{\dag}_{\pm}}{ \sqrt{L}} \exp\left( - \tilde A_{\pm
,i}\left( x  \right) \right) \label{ferm1b}
\end{equation}
with $$
\tilde A^{\dag}_{\pm ,i}\left( x  \right)=
\int_{2\pi/L}^{\infty}\frac{dp}{2\pi}
\sqrt{\frac{2\pi}{ p}}\exp{(\mp ipx)}\frac{1}{\sqrt {2}}\left(\hat C^{\dag }
(\pm p) \pm S^{\dag}_{R,L}(p)\right)
$$
one would obtain a  theory with strong interaction. Different
components of the left- (or right -) particles would interact one
with other via the vertex which is proportional to $V_f^c - v_f$.
Thus, as Fermi-field $\hat\Xi$ diagonalizing  Hamiltonian does not
turn into L or R free electrons when interaction is switched off,
it cannot be interpreted as a renormalized electron. Instead one
should talk about diagonalizing transformation similar to
Bogolubov-Valantin one \cite{V} in the theory of 3D superconductivity.

Just as above one can define the charges of the fermion field
$\hat\Xi_+$. They are equal to $ e^*=\sqrt {2\gamma} e_0$ and
$\sigma=1/\sqrt{2\gamma }$ with $ \gamma = v_f/V^c_f .$

\section{Conclusion}
Using free fermion representation for Luttinger model with
short-range interaction we extend concept of the Fermi liquid
Landau to the system of 1D interacting fermions. The quantum
numbers of Landau quasi-particles are different in comparison with
original electrons due to the strong polarization of the ground
state. In particular, in the limit of infinitely strong
repulsive interaction the electric charge of the quasi-particle is screened
out completely. At the same time chirality of the quasi-particle
in this limit rises steeply, so that their product is equal to
free electron one. These quasi-particles describe all excited
state of the Luttinger liquid and are an analogy with the normal
excitations of Bogolubov-Valantin in the theory of 3D
superconductivity.

\begin{acknowledgments}
We are grateful to V.L.Gurevich for a number of interesting
discussions and reading the manuscript. V.V.A. is grateful for
partial support the work by the Russian National Foundation for
Basic Research, grant No 09-02-00396-A.
\end{acknowledgments}

\appendix
\section{From boson to fermion representation.}
\label{f}

In order to prove that in $\chi$ representation the system is
described by a free Hamiltonian one has to repeat the
bosonization procedure in the reverse order (see, for example, \cite{ML},\cite{P2}, \cite{Hold}).
First of all, we associate the density of diagonalizing
Fermi-fields  ($\hat\varrho_{\pm}$) with the diagonalizing boson
fields $\hat C \left( \pm p\right)$. As usual we define the
density of the exact electron as
\begin{equation}
\hat\varrho_{\pm}\left( x  \right) =\frac{1}{2}\lim_{\epsilon \to 0}
\left(\hat \chi^{\dag}_{\pm}\left( x +\epsilon/2 \right) \hat
\chi_{\pm} \left( x  -\epsilon/2 \right) +\hat
\chi^{\dag}_{\pm}\left( x -\epsilon/2 \right) \hat
\chi_{\pm}\left( x +\epsilon/2\right)\right) .
\label{ferm3}
\end{equation}
We have to use this definition because the quantity $\hat
\chi^{\dag}_{\pm}\left( x \right) \hat \chi_{\pm} \left( y
\right)$ is singular at the point $ x\to y $. The singularity has
to be redefined in such a way that its Fourier transform
$\hat\varrho_{+}\left( p  \right)$ would be determined by $\hat
C\left( p \right)$ for $p > 0$ and $\hat C^{\dag}\left( - p
\right)$ for $p < 0$ in the usual way. Definition (\ref{ferm3})
satisfies this requirement. Indeed, substituting Eq.(\ref{ferm1})
in Eq.(\ref{ferm3}) one gets:
\begin{equation}
\hat\varrho_+\left( x  \right)= \int_{0}^{\infty}\frac{dp}{2\pi}\sqrt{\frac{p}{2\pi}}
\left( \hat C\left( p
\right)\exp{\left(ipx\right)} +  \hat C^{\dag}\left( p
\right)\exp{\left(-ipx\right)} \right) \label{ferm4}
\end{equation}
in accordance with analogous equation for $R$-electron density
(\ref{bos}). Using Eq.(\ref{ferm4}) and a similar one for
$\hat\varrho_-\left( x  \right)$ one can rewrite Hamiltonian
Eq.(\ref{diag}) in the form:

\begin{equation}
H=v_f^c\int^{\infty}_0 d p \int dxdy
\left[\hat\varrho_+\left( x  \right)\hat\varrho_+\left( y  \right)\exp
(ip \left(x - y  \right)) + \hat\varrho_-\left( x
\right)\hat\varrho_-\left( y \right)\exp (ip \left(y - x\right))
\right] \label{diagf1}
\end{equation}
For further calculation it is essential that electron (or
hole) part of the  fermion operator $\hat\chi$ is an operator with
simple analytical properties. For example, $ {\frak
b}_{+}^{\dag}\left( x\right)$ has no singularity in the lower
semiplane of the complex $x$, as one can see from  definition
Eq.~(\ref{psi2}), etc.
Let us discuss first
 the term ${\frak a}^{\dag}_+\left( x\right) {\frak
a}_+\left( x\right){\frak a}^{\dag}_+\left( y\right) {\frak
a}_+\left( y\right)$ which arises
from the  product of two densities $\hat\varrho_+$, where
$\hat\varrho_+={\frak a}^{\dag}_+ {\frak a}_+- {\frak
b}^{\dag}_+{\frak b}_+ + {\frak a}^{\dag}_+{\frak b}^{\dag}_+ +
{\frak b}_+{\frak a}_+$. We present it in the form:
$$
1/2\int dxdy\int^{\infty}_0 d p\left[ {\frak a}^{\dag}_+\left(
x\right) {\frak a}_+\left( x\right){\frak a}^{\dag}_+ \left(
y\right) {\frak a}_+\left( y\right)\exp (ip \left(x - y
+i\delta\right)) + {\frak a}^{\dag}_+\left( y\right) {\frak
a}_+\left( y\right){\frak a}^{\dag}_+\left( x\right) {\frak
a}_+\left( x\right)\exp (ip \left(y - x + i\delta \right)) \right]
.$$ Then, one should exchange two central operators ${\frak
a}_+{\frak a}^{\dag}_+$ in each term using the anticommutator:$$
\{ {\frak a}_+^{\dag}\left( x \right) , {\frak a}_+\left( y
\right)\}  = \frac{1}{2\pi i}\cdot \frac{1}{x-y-i\delta} . $$ The
term with four operators vanishes after integration over $p$ (in
accordance with the Pauli principle and such is the case only for
short range e-e interaction) while  the term with two operators
after the same integration gives
$$ -\frac{1}{2\pi }\int dxdy  {\frak
a}^{\dag}_+\left( x\right)\cdot \frac{1}{(y-x-i\delta)^2} {\frak
a}_+\left( y\right) .$$
  An analytical properties of ${\frak
a}_+\left( y\right)$ allow one to present this term in the form
$$\frac{1}{ i}\int dx {\frak a}^{\dag}_+\left( x\right)
\partial_x {\frak a}_+ \left( x\right),
$$ as it should be in accordance with Eq.(\ref{diagf}).
Each term from (\ref{diagf1}) can be considered in the similar way. As
a result one would obtain the whole Hamiltonian Eq.(\ref{diagf}).

\section{Dzyaloshinskii - Larkin approach and ferminization.}
\label{lar} In the appendix we will discuss relation between
representation of the Luttinger model by free fermions
$\hat\chi_{\pm}$ and Dzyaloshinskii-Larkin  approach. In their
paper \cite{Lar} they proved that RPA approach is exact for the
Luttinger liquid with short range interaction. From the outset it
is clear that free fermions in an RPA bubble and $\hat\chi_{\pm}$
are not the same: they have different velocity, $v_f$ and $v^c_f$
respectively. In order to examine interconnection of the fermions
more deeply it is useful to rederive the  Dzyaloshinskii-Larkin
result in  another way.

The theory with arbitrary electron-electron interaction can be
reduced to a noninteracting fermion theory in an external field by
means of the well-known Hubbard-Stratonovich transformation
\cite{Hub}:
\begin{eqnarray}
\exp  \left[ -\frac{i}{2}\int_{-\infty}^{\infty}dt \int^\infty_{-\infty}
\frac{dp}{2\pi} V \left( p \right) \varrho \left( p,t \right)
\varrho \left( -p,t  \right) \right]
=\protect\phantom{mmmmmmmmmmmmmmm}\nonumber\\
\frac{1}{{\cal N }}\int {\cal D}\Phi \exp \left[ \frac{i}{2}
\int_{-\infty}^{\infty} dt \int^\infty _{-\infty}\frac{dp}{2\pi} \Phi\left( p,t
\right)
\Phi\left( -p,t \right) V^{-1} \left( p  \right)
  -\frac{i}{2} \int_{-\infty}^{\infty} dt \int^\infty
_{-\infty}\frac{dp}{2\pi} \left( \varrho \left( p,t  \right)
\Phi\left( -p,t  \right) + \varrho \left( -p,t  \right) \Phi\left(
p,t  \right)\right) \right]\nonumber
\end{eqnarray}
Here $\cal N$ is normalization coefficient:
\begin{equation}
{\cal N } = \int {\cal D}\Phi \exp \left[  \frac{i}{2} \int_{-\infty}^{\infty} dt
\int^\infty _{-\infty}\frac{dp}{2\pi} \Phi\left( p,t \right)
\Phi\left( -p,t \right) V^{-1} \left( p  \right) \right] \nonumber,
\end{equation}
$V(p)$ is Fourier transformation of e-e interaction, and $\int
{\cal D}\Phi$ is a functional integral over the Bose - fields
$\Phi(x,t)$. The transformation allows one to reduce any electron
Green function $G_{cul}(x_1,x_2,..x_1'..)$ to similar connected
Green function of noninteracting electrons in external electric
field  $\Phi (x,t)$, $G(\Phi ,x_1,x_2,....)$:
\begin{equation}
G_{cul}(x_1,..x_1'..)=\int {\cal D}\Phi G(\Phi ,x_1,..x_1'..)\exp
{\left( S_N(\Phi) + \log {{\rm Det}\hat S(\Phi)} \right)} /\int
{\cal D}\Phi \exp {\left( S_N(\Phi) + \log {{\rm Det}\hat S(\Phi)}
\right)},
\label{b2}
\end{equation}
here $$
S_N(\Phi)= \frac{i}{2}
\int_{-\infty}^{\infty} dt \int^\infty _{-\infty}\frac{dp}{2\pi} \Phi\left( p,t
\right)
\Phi\left( -p,t \right) V^{-1} \left( p  \right)
$$ is the phase arising from Hubbard-Stratonovich transformation (the first term in the RHS
of the transformation) while the operator $\hat S(\Phi)$ is
defined by the kernel of the electron action in an external field,
$\rm{Tr}\left( \overline{\Psi}{\cal S}(\Phi) \Psi\right)$. During
derivation of Eq.(\ref{b2}) we have "calculated" the functional
integral over the electron fields:
\begin{equation}
 \int{\cal D}\Psi{\cal D} \overline{\Psi}{\phantom .} \overline{\Psi }(x_1)...\Psi (x_1')...
\exp{\rm{Tr}\left( \overline{\Psi}{\hat\cal S}(\Phi) \Psi\right)}=
G(\Phi ,x_1,..x_1'..)\exp {\left( \log {{\rm Det}\hat S(\Phi)}
\right)} \label{b3}
\end{equation}
It is important that in order to calculate a Green function one
has to integrate over the fields $\Psi(x,t)$ decaying in the limit
$t\to \pm\infty $. The LHS  of Eq.(\ref{b3}) is the disconnected
electron Green function in an external field (with all electron
loops). So, $\log {{\rm Det}\hat S(\Phi)}$ is the sum of all
connected electron loops in an external field. During the
calculation of the quantity one has to take into account an
ultraviolet divergence. One has to regularize it in a usual way
demanding a gauge-invariance of the theory. For the Luttinger
spinless fermion model it has been calculated in \cite{A} and
equal
\begin{equation}
\log {{\rm Det}\hat S(\Phi)} =  -\frac{1}{4\pi } \int_{-\infty}^{-\infty} dt
dt_1 \int _{-\infty}^\infty \frac{dp}{ 2\pi} \Phi \left( -p,t \right)
\Phi \left( p,t_1  \right) |p| \exp\left[ -i|p|v_f |t-t_1| \right] =
 \label{b4}
\end{equation}
$$
 -\frac{i}{2\pi } \int_{-\infty}^{-\infty}\frac{dpd \omega }{(2\pi)^2 }\Phi (p,\omega)
\Phi (- p,-\omega)\frac{p^2v_f}{\omega^2 - p^2v_f^2 + i\delta}
$$
Notice, the  \eq{b4} is  gauge invariant: a field depending only on time  does not contribute.
So, the phase in Eq.(\ref{b2}) is equal
\begin{equation}
S_N(\Phi) + \log {{\rm Det}\hat S(\Phi)} = \frac{i}{2}
\int_{-\infty}^{-\infty}\frac{dpd \omega }{(2\pi)^2 }\Phi
(p,\omega) \Phi (- p,-\omega)V_0(p)^{-1}\frac{\omega^2 - (p v^{c
}_f)^2+ i\delta}{\omega^2 - p^2v_f^2 + i\delta} \label{b4a}
\end{equation}
In order to check Eqs.(\ref{b2}, \ref{b4a}) let as calculate one
particle Green function. For the case
\begin{equation}
G_{R,L}(\Phi ,x_1,t_1,x_2,t_2)= G_{R,L}^0\left( x_1,t_1;x_2,t_2\right)
\exp \left[ - i \int _{-\infty}^{\infty} dt' \int dy  \Phi (y,t')\left(
G_{R,L}^0\left( x_1,t_1;y,t'\right) - G_{R,L}^0\left( x_2,t_2;y,t'
\right)\right)\right] ,
\label{b5}
\end{equation}where $G_{R,L}^0$ is a free fermion Green function equal to
$$G_{R,L}^0 \left( x_1,t_1;x_2,t_2\right) = \frac{1 } {2\pi i} \left[
v_f \left( t_1-t_2\right) \mp \left( x_1-x_2\right) -i\delta {\rm
sign}\left( t_1-t_2\right) \right] ^{-1} .$$
Thus, linear in $\Phi$ part of the phase in the $G_{R,L}(\Phi ,..)$ can be represent as
$$
\int \frac{d\epsilon dp}{(2\pi )^2} \frac{\exp {(ipx_1-i\epsilon t_1)} -
\exp {(ipx_2-i\epsilon t_2)} } {\epsilon \mp pv_f \pm i\delta {\rm sign}p}\Phi(p,\epsilon ) .
$$
So, Gaussian integral over the fields $\Phi$ can be  easily calculated.
As a result, one has the well-known expression for the short range interaction model:
$$
G_{cul\phantom {.}R}\left( x_1,t_1;x_2,t_2\right) = \frac{1 } {2\pi i} \left[
v_f^c \left( t_1-t_2\right) - \left( x_1-x_2\right) -i\delta {\rm
sign}\left( t_1-t_2\right) \right] ^{-1}\\
\left[(\delta )^2/ (x_1 - x_2)^2 - (v_f^c(t_1-t_2) -i\delta)^2    \right]^\alpha ,
$$ here $\alpha$ is $\sinh^2\theta = (v_f^c-v_f)^2/ 4v_f^cv_f $ while $\theta$
is the diagonalizing angle has been introduced above.

Returning to the our question one can see from (\ref{b4a}) that
effective interaction (the propagator of auxiliary boson field)
equals:
\begin{equation}
V^{eff}(p) = V_0(p)\frac{\omega^2 - p^2v_f^2 }{\omega^2 - (p v^{c
}_f)^2 + i\delta}. \label{b6}
\end{equation}
In \cite{Lar} the same result have been calculated for the short
range interaction ($V_0(p) \to V_0$). It is important that the
result was obtained from the equation for the effective
interaction between $I$ and $J$ electrons, $V_{IJ}$,:
$$ V_{RR} = V_0 + V_0\prod_{R}V_{RR} + V_0\prod_{L}V_{LR}$$
$$V_{LR} =V_0 + V_0\prod_{R}V_{RR} + V_0\prod_{L}V_{LR},$$
where $\prod_{J}$ is fully renormalized electron bubble. Yet, in
\cite{Lar} it was shown that the solution of the equations is the
same as for the idem equations but with free electron bubble
(consisting of free $J-$electron Green functions and without
vertexes renormalization). It means that the RPA approach is exact
for the Luttinger liquid. As it should be, $ V_{RR} $ coincides
with  our $V^{eff}(p)$.

Notice, the limitation of the Dzyaloshinskii - Larkin result for
short range interaction is needless. As one can see from
discussion given above (see Eqs.\ref{b4a}, \ref{b6}) the RPA
approach is valid for arbitrary $V_0(p)$. At the same time the
short range interaction limitation is  crucial for our Fermi
liquid picture. A free fermion description is possible only at the
length larger than characteristic size of an interaction. An
attempt to generalize the proof of Appendix~\ref{f} for arbitrary
interaction brings to four - fermions interaction is forbidden for
the short range case. So, there is no direct relation between
validity of the RPA approach and Fermi liquid description of the
theory. One can give an opposite example. The result (\ref{b6}) is
valid so far as in the effective action the electron loops with a
bigger number of the external fields $\Phi$ do not exist. The last
depends not only of a model but and of a quantity calculated. For
a Green function, the effective action for LL in terms of
$\Phi$-fields  is free. Would one calculate a wave function (of a
ground or an excited state) one has to use a functional integral
defined on a set of the fields $\Psi(x,t)$ non-decaying in time
(for a fermion theory, e.g. for Luttinger liquid, it was shown in
\cite{A}). It gives rise to an additional (exponential in $\Phi$)
terms in the action.  As a result, calculation a functional
integral (pairwise connections of the fields $\Phi$) brings to
strong interaction in a boson system (with vertex renormalization,
etc). It means that a RPA approximation will not exact for the
case.

 \end{document}